\documentclass[aps,english,prb,amsmath,amsfonts,amssymb,superscriptaddress,twocolumn,showpacs,citeautoscript]{revtex4-1}
\usepackage[T1]{fontenc} \usepackage[latin9]{inputenc}
 \usepackage{amsmath} \usepackage{amssymb}
\usepackage{wasysym} \usepackage{graphicx} \usepackage{times}
\usepackage{nicefrac} \usepackage{appendix} \usepackage{textcase}
\usepackage{subfigure} \usepackage{empheq} \usepackage{color}
\usepackage{verbatim}
   
\DeclareMathAlphabet{\mathbfit}{OML}{cmm}{b}{it}

\renewcommand{\Re}{\operatorname{Re}}
\renewcommand{\Im}{\operatorname{Im}}
\renewcommand{\vec}[1]{\mathbf{#1}}

\newcommand{\PhiHT}{\Phi_{12}}
\newcommand{\Phirad}{\Phi_0}
   
\newcommand{\figref}[1]{Fig.~\ref{fig:#1}}
\newcommand{\Figref}[1]{Figure~\ref{fig:#1}}

\renewcommand{\eqref}[1]{(\ref{eq:#1})}

\newcommand{\citeasnoun}[1]{Ref.~\onlinecite{#1}}

\newcommand{\PT}{\mathcal{PT}}
\def\a{s}
\def\b{s}
\newcommand{\add}[1]{\if\a\b{{\color{red} #1}}\else{#1}\fi}
\newcommand{\comm}[1]{\if\a\b{{\color{blue}\{\small \sc #1\}}}\else{}\fi}
\newcommand{\del}[1]{{\if\a\b{{\color{magenta}[[#1]]}}\else{}\fi}}

\begin{document}

\title{Giant frequency-selective near-field energy transfer in
  active--passive structures}
 
\author{Chinmay Khandekar}
\affiliation{Department of Electrical Engineering, Princeton University, Princeton, NJ 08544, USA}
\author{Weiliang Jin}
\affiliation{Department of Electrical Engineering, Princeton University, Princeton, NJ 08544, USA}
\author{Owen D. Miller}
\affiliation{Department of Mathematics, Massachusetts Institute of Technology, Cambridge, MA 02139, USA}
\author{Adi Pick}
\affiliation{Department of Physics, Harvard University, Cambridge, MA 02138, USA}
\author{Alejandro W. Rodriguez}
\affiliation{Department of Electrical Engineering, Princeton University, Princeton, NJ 08544, USA}

\begin{abstract}
  We apply a fluctuation electrodynamics framework in combination with
  semi-analytical (dipolar) approximations to study amplified
  spontaneous energy transfer (ASET) between active and passive
  bodies. We consider near-field energy transfer between semi-infinite
  planar media and spherical structures (dimers and lattices) subject
  to gain, and show that the combination of loss compensation and
  near-field enhancement (achieved by the proximity, enhanced
  interactions, and tuning of subwavelength resonances) in these
  structures can result in orders of magnitude ASET enhancements below
  the lasing threshold. We examine various possible geometric
  configurations, including realistic materials, and describe optimal
  conditions for enhancing ASET, showing that the latter depends
  sensitively on both geometry and gain, enabling efficient and
  tunable gain-assisted energy extraction from structured surfaces.
\end{abstract} 

\maketitle

%Such an enhancement provides opportunities for applications in energy
%and power generation~\cite{Laroche06:prl}, thermal
%imaging~\cite{Satoshi09}, nanoscale local heat
%management~\cite{Segal08,Nitzan06}, thermal
%rectification~\cite{Biehs13,Otey10} and radiative
%cooling~\cite{heatfluxFan15}. Recently, there is increased interest in
%understanding ways to control and extract heat from active
%structures~\cite{Chen14,Minnich15,heatfluxFan15}.

Radiative heat transfer between nearby objects can be much larger in
the near field (sub-micron separations) than in the far
field~\cite{Whiting11,Ilic12,Francoeur08} due to coupling between
evanescent (surface-localized) waves~\cite{BasuZhang09,Loomis94}.  In
this paper, we investigate the possibility of exploiting both active
materials and geometry to enhance and tune near-field energy
transfer. In particular, we study amplified spontaneous energy
transfer (ASET)---the amplified spontaneous emission (ASE) from a gain
medium that is absorbed by a nearby passive object---and demonstrate
orders of magnitude enhancements compared to far-field emission or
transfer between passive structures. Our work extends previous work on
heat transfer between planar, passive
media~\cite{Biehs11:apl,Ben-Abdallah09,Narayanaswamy08:spheres,
  narayanaswamy2008near} to consider the possibility of using gain as
a mechanism of loss cancellation, leading to further flux-rate
enhancements under certain conditions (diverging at the onset of
lasing). Since planar structures are known to be sub-optimal
near-field energy transmitters~\cite{miller2016fundamental}, we also
consider a more complicated geometry involving subwavelength metallic
dimers or lattices of spheres doped with active emitters, and describe
conditions under which ASET $\gg$ ASE below the lasing threshold
(LT). Our analysis of these spherical structures includes both
semi-analytical calculations (for dimers) and dipolar approximations
that include first-order geometric modifications to the polarization
response of spheres (for lattices), revealing not only significant
potential enhancements but also strongly geometry-dependent variations
in ASET stemming from the presence of multiple scattering, which
suggests the possibility of using the near field as a mechanism for
tuning energy extraction. Similar to our recent findings in the case
of passive objects~\cite{miller2015shape}, we find that energy
exchange between lattice of spheres tends to greatly outperform
exchange between planar bodies as the intrinsic loss rates of
materials decrease, with gain contributing additional enhancement.

% We further analyze the complex relationship between heat transfer,
% far-field emission and dimer parameters, and find that in contrast to
% heat transfer between passive objects, limits to ASET are not uniquely
% optimal, occurring for multiple configurations of sphere radii,
% permittivities and separations, in contrast to heat transfer between
% passive objects which is often a monotonic function of object
% separations.
 
Recent approaches to tailoring incoherent emission from nanostructured
surfaces have begun to explore situations that deviate from the usual
linear and passive
materials~\cite{Guo09,Wenzel96,Christ11,Yang13,chinmay15,chinmay15:pumped},
with the majority of these works primarily focusing on ways to control
far-field emission, e.g. the lasing properties of active
materials~\cite{Peng14:science}. Here we consider a different subset
of such systems: structured active--passive bodies that exchange
energy among one another more efficiently than they do into the far
field. Our predictions below extend recent progress in understanding
and tailoring energy exchange between structured materials, which thus
far include doped semiconductors~\cite{Zhang14}, phase-change
materials~\cite{Biehs13,Wang13}, and metallic
gratings~\cite{Yamada15,Rodriguez11:PC,Song15}. Active control of near
field heat exchange offers a growing number of applications, from heat
flux control~\cite{Chen14,heatfluxFan15} and solid-state
cooling~\cite{heatfluxFan15} to thermal
diodes~\cite{Chevrier12,Fan12}. Our work extends these recent ideas to
situations involving systems undergoing gain-induced amplification.

The starting point of our analysis is the well-known linear
fluctuational electrodynamics framework established by Rytov, Polder,
and van~Hove~\cite{Rytov89,PolderVanHove71}. In particular, given two
bodies held at temperatures $T_1$ and $T_2$, and separated by a
distance $d$, the power or heat transfer from $1 \to 2$ is given
by~\cite{BasuZhang09}:
\begin{align}
  P(T_1,T_2)=\int_0^{\infty} [\Theta(\omega,T_1)-\Theta(\omega,T_2)]
  \PhiHT(\omega) \frac{d\omega}{2\pi}
\label{eq:flux}
\end{align}
where $\Theta(\omega,T)$ is the mean energy of a Planck oscillator at
frequency $\omega$ and temperature $T$, and $\PhiHT(\omega)$ denotes
the spectral radiative heat flux, or the absorbed power in object 2
due to spatially incoherent dipole currents in 1. Such an expression
is often derived by application of the fluctuation-dissipation theorem
(FDT), which relates the spectral density of current fluctuations in
the system to dissipation~\cite{BasuZhang09}, $\langle
J_i(\vec{x},\omega),J_j^*(\vec{x}',\omega)\rangle=\frac{4}{\pi}\omega\epsilon_0
\Im\epsilon(\vec{x},\omega)
\delta(\vec{x}-\vec{x}')\Theta(\omega,T)\delta_{ij}$, where $J_i$
denotes the current density in the $i$th direction, $\epsilon_0$ and
$\epsilon(\vec{x}, \omega)$ are the vacuum and relative permittivities
at $\vec{x}$, and $\langle \cdots \rangle$ denotes a thermodynamic
ensemble-average.

Extensions of the FDT above to situations involving active media
require macroscopic descriptions of their dielectric response. Below,
we consider an atomically doped gain medium that, ignoring stimulated
emission or nonlinear effects arising near the lasing
threshold~\cite{Matloob97}, can be accurately modelled (under the
stationary-inversion approximation) by a simple two-level Lorentzian
gain profile of the atomic populations $n_1$ and $n_2$, resulting in
the following effective permittivity~\cite{Vaillon10}:
\begin{align}
  \epsilon(\omega)=\epsilon_r(\omega) + \underbrace{\frac{4\pi
      g^2}{\hbar
      \gamma_{\bot}}\frac{\gamma_{\bot}D_0}{\omega-\omega_{21}+i\gamma_{\bot}}}_{\epsilon_G(\omega)}
\label{eq:gaindispersion}
\end{align}
where $\epsilon_r$ denotes the permittivity of the background medium
and the second term describes the gain profile $\epsilon_G$, which
depends on the ``lasing'' frequency $\omega_{21}$, polarization decay
rate $\gamma_{\bot}$, coupling strength $g$, and population inversion
$D_0=n_2-n_1$ associated with the $2 \to 1$ transition.
Detailed-balance and thermodynamic considerations lead to a modified
version of the FDT~\cite{Matloob97,graham68,jeffers93} involving an
effective Planck distribution $\Theta(\omega_{21},T_G) =
-n_2\hbar\omega_{21}/D_0$, in which case the system exhibits a
negative effective or ``dynamic'' temperature under $n_2 >
n_1$~\cite{jeffers93}. Note that even though $\Theta < 0$ under
population inversion, the radiative flux from such a medium is
positive-definitive: because $\Im \epsilon_G < 0$, the spectral
electric-current correlation function associated with the active
medium,
\begin{equation}
\langle J_i(\vec{x},\omega) J^*_j(\vec{x}',\omega) \rangle =
-\frac{4}{\pi}\omega\epsilon_0(\Im\epsilon_G) \underbrace{n_2\hbar\omega_{21}/D_0}_{\Theta(\omega_{21},T_G)}\delta(\bf{x}-\bf{x}')\delta_{ij}
\end{equation}
is positive. As a consequence, the heat transfer originating
from atomic fluctuations in an active body to a passive body always
flows from the former to the latter, i.e. $T<0$ reservoirs always
transfer energy~\cite{Matloob97}. Of course, in addition to
fluctuations of the polarization of the gain atoms, such a medium will
also exhibit fluctuations in the polarization of the host medium,
depending on its thermodynamic temperature and background loss rate
$\sim \Im\varepsilon_r$, as described by the standard
FDT~\cite{BasuZhang09}. Although thermal flux rates can themselves be
altered (e.g. enhanced) in the presence of gain through the dependence
of $\Phi_{12}$ on the overall permittivity, the flux rate from such an
active medium will tend to be dominated by the fluctuations of the
gain atoms, the focus of our work.

\begin{figure}[t!]
\centering
\includegraphics[width=0.6\linewidth]{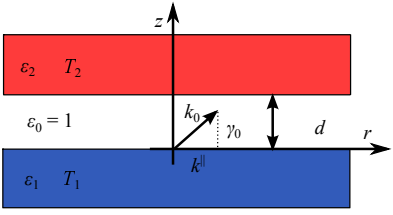}
\caption{Schematic of two semi-infinite plates of permittivities
  $\epsilon_1$ and $\epsilon_2$, respectively, separated by a vacuum
  gap $d$. Fourier decomposition of scattered waves with respect to
  parallel $k_\parallel$ and perpendicular $\gamma$ wavevectors
  simplifies calculations of energy transfer.}
\label{fig:plates}
\end{figure}

\section{Planar media}

We begin our analysis of ASET by first considering an extensively
studied geometry involving two semi-infinite plates that exchange
energy in the near field. Such a situation has been thoroughly studied
in the past in various
contexts~\cite{Biehs11:apl,Ben-Abdallah09,Narayanaswamy08:spheres,
  narayanaswamy2008near}, but with passive materials, whereas below we
consider the possibility of optical gain in one of the plates. For
simplicity, we omit the frequency dependence in the complex dielectric
functions $\epsilon_j$ of the two plates ($j=1,2$), shown
schematically in \figref{plates} along with our chosen coordinate
convention. We assume that one of the plates is doped with a gain
medium, such that $\epsilon_1=\epsilon_r+\epsilon_G$, and consider
only fluxes due to fluctuations in the active constituents $\sim \Im
\epsilon_G$, as described by the modified FDT
above~\cite{BasuZhang09,pick2015multimode}. Due to the translational
symmetry of the system, it is natural to express the heat flux in a
Fourier basis of propagating transverse waves
$k_\parallel$~\cite{BasuZhang09}, in which case the flux is given by
an integral $\Phi(\omega)=\int \Phi(\omega,k_\parallel) k_\parallel
dk_\parallel$. In the near field, $k_\parallel > \omega/c$, the main
contributions to the integrand come from evanescent waves which
exchange energy at a rate~\cite{Rytov89,Loomis94},
\begin{align}
  \PhiHT(\omega,k_\parallel) \approx \sum_{q=s,p}
  \frac{\Im(\epsilon_{G})\Im(r_{1}^q)\Im(r_{2}^q)
    e^{-2\Im(\gamma_0)d}}{\Im \epsilon_1\left|1-r_{1}^q r_{2}^q
    e^{-2\Im(\gamma_0)d}\right|^2},
\label{eq:PhiPP}
\end{align}
where $r^s_{j}=\frac{\gamma_0-\gamma_j}{\gamma_0+\gamma_j}$ and
$r^p_{j}=\frac{\epsilon_j\gamma_0 -\gamma_j}{\epsilon_j\gamma_0 +
  \gamma_j}$ are the Fresnel reflection coefficients at the interface
between vacuum and the dielectric media, for $s$ and $p$
polarizations, respectively, defined in terms of the wavevectors
$\vec{k}_j = k_\parallel \mathbf{\hat{r}} + \gamma_j
\mathbf{\hat{z}}$, with $|\vec{k}_0|=\omega/c$ and
$|\vec{k}_j|^2=k_\parallel^2+\gamma_j^2 =
\epsilon_j\omega^2/c^2$. Note that the derivation of Fresnel
coefficients requires special care since when gain compensates loss,
i.e.  $\Im\epsilon_1<0$, the sign of the perpendicular wavevector
$\gamma_1=\pm\sqrt{\epsilon_1\omega^2/c^2-k_\parallel^2}$ needs to be
chosen correctly inside the gain
medium~\cite{Skaar06,kinsler2009refractive,nistad2008causality}. Here,
we make the physically motivated choice that yields decaying surface
waves inside the semi-infinite gain medium. In the case of evanescent
waves $k_\parallel \gg \omega/c$, $\gamma_0 \approx \gamma_j \approx
ik_\parallel$, such that $r^s_j\rightarrow 0$ and
$r^p_j=\frac{\epsilon_j-1}{\epsilon_j+1}=
\frac{|\epsilon_j|^2-1}{|\epsilon_j+1|^2}+
\frac{2\epsilon_j''i}{|\epsilon_j+1|^2}$, where
$\epsilon_j=\epsilon_j'+i\epsilon_j''$. Substituting $e^{2k_\parallel
  d}=z$ and approximating the integral $\int z f(z) dz \approx z_0
f(z)$, with $z_0 = k_0 d=\ln{|r^p_1 r^p_2|}$ denoting the wavevector
that minimizes the denominator of \eqref{PhiPP}, one obtains:
\begin{multline}
\Phi_{12}(\omega) =\frac{z_0 \Im(\epsilon_G)\Im(r^p_1)\Im(r^p_2)}{4\pi^2
  d^2\Im\epsilon_1} \\ \times \int_1^\infty \frac{dz}{(z-\Re(r^p_1
  r^p_2))^2+(\Im(r^p_1 r^p_2))^2}
\label{eq:phi12plates}
\end{multline}
It follows that the flux rate in the case of passive media with small
loss rates scales as $\Phi_{12}\approx \ln|r^p_1 r^p_2|/ ( 4\pi^2 d^2)
\sim \frac{1}{d^2}\ln
|\frac{\epsilon_1-1}{\Im\epsilon_1}\frac{\epsilon_2-1}{\Im\epsilon_2}|$
under the resonant condition $\Re\epsilon_j=-1$, illustrating a slow,
logarithmic dependence on the loss rates and corresponding divergence
as $\Im\epsilon_j \to 0$, described
in~\citeasnoun{miller2015shape}. However, ASET in the presence of
gain, described by \eqref{phi12plates}, depends differently on the
loss rates. On the one hand, in situations where gain does not
compensate for losses ($\Im\epsilon_1>0$), the integral can be further
simplified to yield $\Phi_{12}\approx
\frac{1}{d^2}\frac{\Im\epsilon_G}{\Im\epsilon_1}
\ln|\frac{\epsilon_1-1}{\Im\epsilon_1}\frac{\epsilon_2-1}{\Im\epsilon_2}|$,
illustrating the same logarithmic dependence on loss rates and
resonant conditions, but with the flux rate exhibiting an additional
factor $\sim \Im\epsilon_G/\Im \epsilon_1$. On the other hand, when
the active plate has overall gain, i.e. $\Im\epsilon_1 <0$, the
integral diverges under the modified condition $\Re(r^p_1 r^p_2)>1$
and $\Im(r^p_1 r^p_2)=0$, or alternatively,
\begin{align}
&(|\epsilon_1|^2-1)(|\epsilon_2|^2-1)-4\epsilon_1''\epsilon_2'' >
  |\epsilon_1+1|^2|\epsilon_2+1|^2 \\
&\epsilon_2''(|\epsilon_1|^2-1)+\epsilon_1''(|\epsilon_2|^2-1)=0
\end{align}
both of which cannot be simultaneously satisfied below threshold. Note
that in this regime, $\Re\epsilon=-1$ is no longer a necessary
condition for maximum heat transfer. In particular, the divergence can
occur at unequal values of $\Re\epsilon_j$ and $\Im\epsilon_j$, in
which case the linewidth $\sim |\Im (r_1^p r_2^p)|$ and peak
wavevector $\sim \Re (r_1^p r_2^p)$ are decreased and increased,
respectively, by suitable choices of material parameters. Such a
divergence is of course indicative of a LT, at which point linear
fluctuational electrodynamics is no longer valid. Although
semi-infinite plates offer analytical insights and computational ease,
their closed nature and large effective loss rates make them far from
ideal for studying ASET. In what follows, we consider finite and open
geometries in which even larger ASET and tunability can be attained.

\section{Sphere Dimers and Lattices}

\begin{figure}[t!]
\centering
\includegraphics[width=0.6\linewidth]{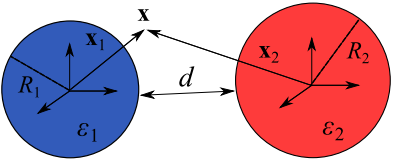}
\caption{Schematic of dimer system consisting of two spheres of
  permittivities $\epsilon_1$ and $\epsilon_2$ and radii $R_1$ and
  $R_2$, respectively, and separated by a gap $d$. Mie-series
  decomposition of scattered fields simplifies calculations of energy
  transfer; shown are a flux evaluation point
  $\vec{x}=\vec{x}_1=\vec{x}_2$ in medium 0, with $\vec{x}_{i}$,
  denoting the position relative to the center of sphere $i$.}
\label{fig:spheres}
\end{figure}

\subsection{Sphere dimers}

Consider an illustrative open geometry consisting of two spheres
separated by vacuum, shown in \figref{spheres}. In addition to
material loss, such a system also suffers from radiative losses, which
we quantify (neglecting stimulated emission) from the far-field flux
$\Phirad$. The calculation of heat transfer between two spheres was
only recently carried out using both
semi-analytical~\cite{Narayanaswamy08:spheres} and brute-force
methods~\cite{Rodriguez15:fvc}. Here, we extend these studies to
consider far-field radiation from one of the spheres (in the presence
of the other) and the possibility of gain.  In particular, we analyze
near-field energy exchange $\Phi_{12}$ and far field emission
$\Phi_0$ by exploiting a semi-analytical method (SA) based on
Mie-series expansion of scattered waves, and which follows from a
recent study of heat transfer in a similar but passive
geometry~\cite{Narayanaswamy08:spheres}.

Due to the spherical symmetry of each object, it is natural to
consider scattering in this system by employing field expansions in
terms of Mie series~\cite{Huffman98}. \Figref{spheres} shows a
schematic of the system, consisting of two vacuum-separated spheres of
radii $R_j$ and dielectric permittivities $\epsilon_j$, separated by
surface--surface distance $d$, where one of the spheres is doped with
a gain medium, such that $\epsilon_1=\epsilon_r+\epsilon_G$. We
compute the flux rates through a surface $S$ in vacuum from dipoles
$\vec{x}'_1 \in V_1$ which is given by $\Re \oint_S \langle
\vec{E}^*\times\vec{H}\rangle = \frac{\omega^2 \Im \epsilon_G}{\pi}
\Im \oint_S \int_{V_1} d^3\vec{x}_1'\, \mathbb{G}^* \times (\nabla
\times \mathbb{G}) \cdot d\vec{S}$, where
$\mathbb{G}(\vec{x},\vec{x}_1')$ is the Dyadic Green's function (GF),
or the electric field due to a dipole source at $\vec{x}_1'$ evaluated
at a point $\vec{x}=\vec{x}_1=\vec{x}_2$ in vacuum, with $\vec{x}_{j}$
denoting the position relative to the center of sphere $j$, and where
we have employed the FDT above to express the flux as a sum of
contributions from individual (spatially uncorrelated) dipoles.
 
When expressed in a basis of Mie modes, the GF from a dipole at a
position $\vec{x}_1'\in V_1$ evaluated at $\vec{x}$ is given
by~\cite{Narayanaswamy08:spheres}:
\begin{multline}
%\begin{widetext} \begin{equation}
  \mathbb{G}(\vec{x},\vec{x}'_1) = i k_0
  \sum_{\substack{\ell,\nu=(1,m) \\m=-N}}^{\substack{\ell,\nu=N
      \\m=N}} (-1)^m \sum_{\substack{q,q'=\pm}}
  \vec{M}^{(1)q'}_{\ell,-m}(k_1\vec{x}'_1) \,\otimes \\ \left[C_{\nu m}^{\ell q
      q'} \vec{M}^{(3)q}_{\nu m}(k_0\vec{x}_1) + D_{\nu m}^{\ell q q'}
    \vec{M}^{(3)q}_{\nu m}(k_0\vec{x}_2)\right],
%\label{eq:G}
%\end{equation}
%\end{widetext}
\end{multline}
where $k_j = \sqrt{\epsilon_j} \omega/c$, $\ell \in \mathbb{Z}^+$,
$|m| \leq \ell$, $N$ denotes the maximum Mie order, $C^{\ell q
  q'}_{\nu m}$ and $D^{\ell q q'}_{\nu m}$ are standard Mie
coefficients~\cite{Huffman98,Wang93}, $\vec{M}^{(p)\pm}_{\ell m}$
denote spherical vector waves, $z^{(p)}_\ell$ are spherical Bessel
($p=1$) and Hankel ($p=3$) functions of order $\ell$,
$\zeta^{(p)}_\ell(x) = \frac{1}{x}\frac{d}{dx} [x z^{(p)}_\ell(x)]$,
and $\vec{V}^{(p)}_{\ell m}$ are spherical vector
harmonics~\cite{Chew99}.

The advantages of employing spherical vector waves comes from the
useful orthogonality relations~\cite{Narayanaswamy08:spheres}
described in Appendix \ref{appendix:app}, which greatly simplify the
calculation of fluxes, requiring integration over $V_1$ and over
either the surface $S: |\vec{x}_2| \to R_2$ circumscribing sphere 2
(as derived previously in~\citeasnoun{Narayanaswamy08:spheres}) or a
far-away surface $S: |\vec{x}| \to \infty$, leading to the following
expressions:
\begin{align}
    \label{eq:PhiNF}
    \PhiHT(\omega) &= \frac{R_1\Im\epsilon_G}{R_2\Im\epsilon_1}
    \sum_{\substack{m,\ell,\nu \\ q,p=\pm}}
    \Im\left(\frac{1}{x^{q}_\nu(R_2)}\right)
    \Im\left(\frac{1}{x^{p}_\ell(R_1)}\right) \nonumber
    \\ &\hspace{0.7in} \times \left|\frac{z^{(1)}_\ell(k_1 R_1)
      D^{\ell q p}_{\nu m}}{z^{(1)}_\nu(k_0 R_2)}\right|^2
    |x^{p}_\ell(R_2)|^2, \\ \Phirad(\omega) &= \frac{2k_0^3
      R_1^2\Im\epsilon_G}{\pi\Im\epsilon_1} \sum_{\substack{m,l,\nu
        \\ q,p=\pm}} y^{p}_\ell(R_1) \Big(|D^{\ell q p}_{\nu
      m}|^2+|C^{\ell q p}_{\nu m}|^2 \Big),
    \label{eq:PhiFF}
\end{align}
where $C^{\ell q q'}_{\nu m}$ and $D^{\ell q q'}_{\nu m}$ are
so-called Mie coefficients~\cite{Huffman98},
\begin{align*}
  x^+_\nu(r)&=k_0 r \zeta^{(1)}_\nu(k_1 r) z^{(1)}_\nu(k_0 r)-k_1
  r\zeta^{(1)}_\nu(k_0 r) z^{(1)}_\nu(k_1 r) \\ y^+_\nu(r)&=\lim_{R
    \to \infty} R^2 \Im [z^{(3)}_\nu(k_0 R)\zeta^{(3)*}_\nu(k_0 R)]
  \\ &\hspace{1in} \times \Im [z^{(1)}_\nu(k_1 r)
    \zeta^{(1)*}_\nu(k_1 r)],
\end{align*}
$x^-_\nu(r) = x^+_\nu(r | \zeta \leftrightarrow z)$, $y^-_\nu(r) =
y^+_\nu(r| \zeta \leftrightarrow z)$, $z^{(p)}_\ell$ are spherical
Bessel ($p=1$) and Hankel ($p=3$) functions of order $\ell$,
$\zeta^{(p)}_\ell(x) = \frac{1}{x}\frac{d}{dx} [x z^{(p)}_\ell(x)]$,
and $k_j=\omega\sqrt{\epsilon_j}/c$. We note that \eqref{PhiFF}
appears to be new, but we have checked its validity against
numerics~\cite{Rodriguez15:fvc} and also known expressions in the
limit ($d\to\infty$) of an isolated sphere~\cite{Huffman98}. We also
note that the factors of $\Im\epsilon_G/\Im\epsilon_1$ in both flux
expressions arise because we only consider fluctuations arising from
the active constituents (same as in Eqs.\eqref{PhiPP} and
\eqref{phi12plates} for plates).

%\begin{widetext}
%TODO: A note on our notation above for the Mie coefficients and their
%calculation.

\begin{figure}[t!]
\centering 
\includegraphics[width=1\linewidth]{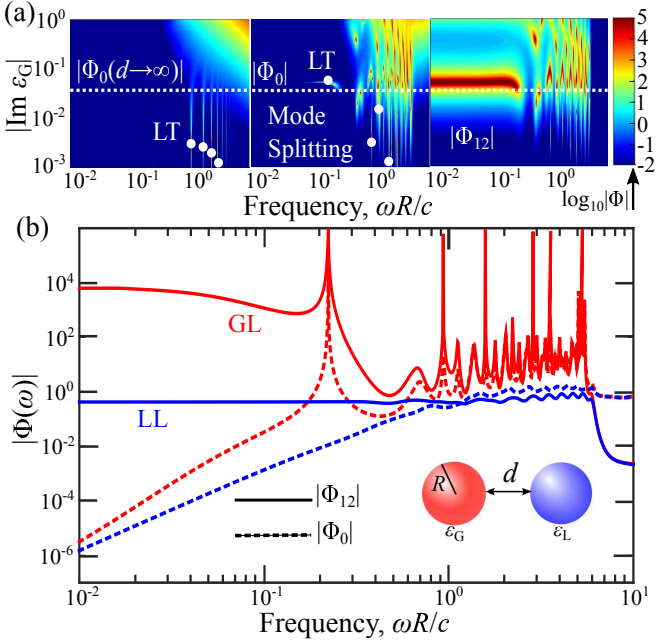}
\caption{Far-field flux $\Phirad(\omega)$ and flux-transfer
  $\PhiHT(\omega)$ associated with a dimer of two spheres of equal
  radii $R$, permittivities $\epsilon_1=\epsilon_r+\epsilon_G$ and
  $\epsilon_2=\epsilon_r$, with $\Im \epsilon_r=0.05$, and separated
  by distance of separation $d$, under various operating
  conditions. (a) Dependence of $\Phirad(\omega)$ and $\PhiHT(\omega)$
  on $\Im\epsilon_1 < 0$ (under gain) at fixed
  $\Re\epsilon_{1,2}=-1.522$, and for either $d\to \infty$ (left) or
  $d/R=0.3$ (middle/right). White circles indicate the lasing
  threshold of a few individual modes while white dashed lines
  indicate operating parameters (cross sections) for the plots in (b),
  which show $\PhiHT$ (solid lines) and $\Phirad$ (dashed lines) at
  fixed $\Im\epsilon_1=-\Im\epsilon_2=-0.05$ and $d/R=0.3$. The plots
  compare the flux rates of gain-loss (GL) dimers (red lines) against
  those of passive (LL) dimers (blue lines).}
\label{fig:fig1}
\end{figure}

We begin by describing a few of the most relevant radiative features
associated with this geometry, focusing on dimers comprising spheres
of constant (dispersionless) dielectric permittivities
$\epsilon_{1,2}$ and equal radii $R$, which very clearly delineate the
operating conditions needed to observe $\PhiHT \gg \Phirad$. We assume
that one of the spheres (with dielectric $\epsilon_1$) is doped with a
gain medium such that $\Im\epsilon_1 < 0$. The top contour in
\figref{fig1}(a) shows $\Phirad$ from an isolated sphere of
$\Re\epsilon=-1.522$ as a function of gain permittivity
$\Im\epsilon_1$, illustrating the appearance of Mie resonances and
consequently, ASE peaks occurring at $k_0 R \gtrsim 1$. As expected,
the LTs (white circles indicate a select few) associated with each
resonance occur at those values of gain where (as in the planar case)
$\Phirad \to \infty$ and the mode bandwidths $\to 0$, decreasing with
increasing $k_0 R $ (smaller radiative losses). Note that these
divergences are obscured in the contour plot by our finite numerical
resolution, which sets an upper bound on $\Phirad$.  The middle
contour plot in \figref{fig1}(a) shows that a passive sphere with
$\Im\epsilon_2=0.05$ in proximity to the gain sphere ($d/R=0.3$)
causes the Mie resonances to couple and split, leading to dramatic
changes in the corresponding LTs. Noticeably, while the presence of
the lossy sphere introduces additional dissipative channels, in some
cases it can nevertheless enhance ASE (decreasing LTs) by suppressing
radiative losses~\cite{Valle15}. These results are well-studied in the
literature~\cite{Valle15,Peng14:science} but they are important here
because our linear FDT is only valid below LT. Another feature
associated with such dimers is the significant enhancement in $\PhiHT$
compared to $\Phirad$ in the subwavelength regime $k_0 R\ll
1$~\cite{Greffet05,Basu11}, illustrated by the middle/right contours
of \figref{fig1}(a). Although such near-field enhancements have
been studied extensively in the context of passive
bodies~\cite{BasuZhang09,Ben-Abdallah09,Basu11}, as we show here, the
introduction of gain can lead to even further enhancements. This is
demonstrated by the flux spectra in \figref{fig1}(b) (corresponding to
slices of the contour maps, denoted by white dashed lines), which
compare the flux rates of both active (red lines) and passive (blue
lines) dimers. The spectra indicate that, while the large radiative
components of Mie resonances at intermediate and large frequencies
$k_0 R \gtrsim 1$ lead to roughly equal enhancements in $\PhiHT$ and
$\Phirad \sim \PhiHT$, the saturating and dominant contribution of
evanescent fields and the presence of surface--plasmon resonances in
the long wavelength regime cause $\Phirad \to 0$ and $\PhiHT \gg 1$ as
$\omega \to 0$. As expected, the existence and coupling of these
resonances depend sensitively on $d/R$, occurring at $\Re\epsilon
\approx \{-2,-1\}$ in the limit $d \to \{0,\infty\}$ of two
semi-infinite plates or isolated spheres, respectively.

\subsection{Dipolar approximation}

Since $\Phi_{12} \gg \Phi_0$ in the subwavelength regime, we consider
a simple dipolar approximation
(DA)~\cite{Chapuis08apl,joulain2005surface} or quasistatic analysis to
understand these enhancements in more detail. In the quasistatic
regime, treating the spheres as point dipoles, we find that the flux
rates are given by:
\begin{align}
\label{eq:DAPhiNF}
\PhiHT &= \frac{12\Im\epsilon_G}{\pi
  L^6\Im\epsilon_1}\Im\alpha_1^{\text{eff}} \Im\alpha_2^{\text{eff}}
\\ \Phirad &= \frac{4\Im\epsilon_G}{\pi\Im\epsilon_1} (k_0 R)^3 \Im
\alpha_1^{\text{eff}},
\label{eq:DAPhirad}
\end{align} 
where $\alpha^\text{eff}_i$ denote each spheres' effective
\emph{anisotropic} polarizability (computed by taking into account
induced polarization of the dipoles), with parallel ($\parallel$) and
perpendicular ($\perp$) components given by~\cite{Schatz05}:
\begin{align}
  \alpha_{\perp,
    1/2}^{\text{eff}}=\alpha_{1/2}\frac{1-\frac{\alpha_{2/1}}{L^3}}{1-\frac{
      \alpha_1\alpha_2}{L^6}}, \,\,\,\, \alpha_{\parallel,
    1/2}^{\text{eff}}=\alpha_{1/2}\frac{1+\frac{2\alpha_{2/1}}{L^3}}{1
    -\frac{4\alpha_1\alpha_2}{L^6}}
\end{align}
with $\alpha_i=\frac{\epsilon_i-1}{\epsilon_i+2}$ denoting the vacuum
polarizability of the isolated spheres in units of $4\pi R^3$ and
$L=2+\frac{d}{R}$ their center-center distance in units of $R$.

\begin{figure}[t!]
\centering
\includegraphics[width=1\linewidth]{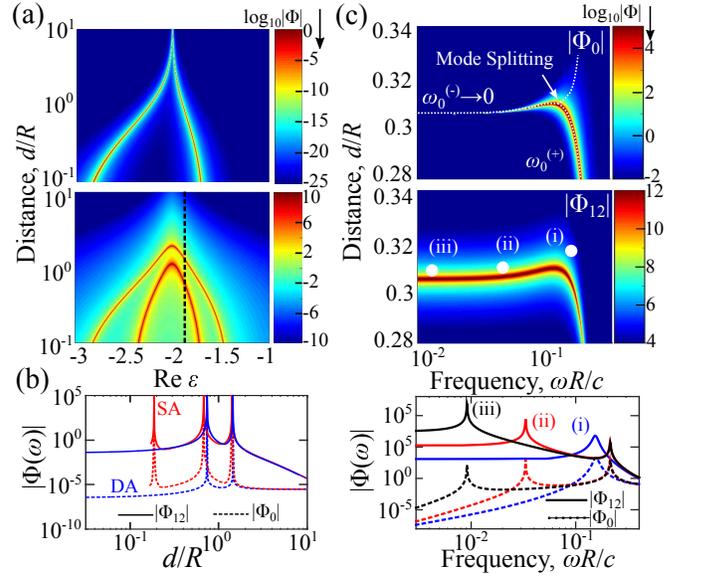}
\caption{(a) Flux-transfer rate $\PhiHT$ associated with the sphere
  dimer system of \figref{fig1} under a simple dipolar approximation
  (DA), in either passive ($\Im\epsilon_{1,2}=0.01$, top) or active
  ($\Im\epsilon_{1}=-\Im\epsilon_2=-0.1$, bottom) regimes, as a
  function of $\Re\epsilon_{1,2}$ and $d/R$. While the flux rate
  diverges in the active case under total loss compensation, only the
  rate per unit volume diverges in the case of finite, passive
  spheres. The validity of the DA for large $d > R$ is illustrated in
  (b), which shows also results obtained using the semi-analytical
  (SA) equations [\eqref{PhiNF} and \eqref{PhiFF}]. (c) Flux rate
  spectra $\Phi_0(\omega)$ (top) and $\PhiHT(\omega)$ (bottom) of the
  dimer system under the $\PT$ symmetry condition,
  $\Re\epsilon_{1,2}=-1.522$ and
  $\Im\epsilon_{1}=-\Im\epsilon_2=-0.05$, illustrating the splitting
  of a sub-wavelength dimer mode as $d$ changes around a critical $d_c
  \approx 0.306R$. The two branches include both quasistatic
  $\omega_0^{(-)}$ and subwavelength $\omega_0^{(+)}$ resonances. (d)
  Flux spectra at three different separations $d \approx
  \{0.3056,0.302,0.3017\}R$, marked by the white dots (i), (ii), and
  (iii), respectively, in the bottom contour in (c).}
\label{fig:fig2}
\end{figure}

It is well known that in the far-field dipolar limit $d/R \gg 1$, both
$\PhiHT, \Phirad \to \infty$ under the resonance condition,
$\Re\epsilon=-2$ and zero material loss $\Im\epsilon \to
0$~\cite{miller2016fundamental,Chapuis08apl,Greffet05}. At smaller
separations, these two conditions are modified to
$|L^6-\alpha_1\alpha_2|=0$ ($\parallel$ component) or
$|L^6-4\alpha_1\alpha_2|=0$ ($\perp$ component) due to changes in the
effective polarizability of each sphere. Despite such a modification,
in the case of passive dimers, the divergence can only be reached in
the limit $\Im \epsilon_i \to 0$. For instance, in passive dimers with
$\alpha=\alpha_1=\alpha_2$, $\Im\alpha^{\text{eff}}\to \infty$ at
specific $L^3=-\Re\alpha$ ($\perp$ component) and $L^3=2\Re\alpha$
($\parallel$ component) for $\Re\epsilon$ close to $-2$ but only under
the condition of zero loss, illustrated in the top contour of
\figref{fig2}(a) for a small $\Im\epsilon_{1,2}=0.01$. Ultimately,
however, the zero-loss quasistatic condition cannot generally be
satisfied in finite, passive geometries, resulting in finite flux
rates (even in the limit as $\Im\epsilon \to 0$); essentially, two
far-separated ($d \to \infty$) spheres will not behave as quasistatic
dipoles owing to their finite skin-depth, except in the limit $R\to 0$
in which case only the flux rates per unit volume rather than the
absolute rates
diverge~\cite{miller2016fundamental,Zhang07}. Gain--loss dimers, on
the other hand, exhibit diverging flux rates (i.e. they can lase)
under finite material gain and loss rates, as well as in finite
geometries that lie outside of the quasistatic regime. A clear and
practical example are objects satisfying the so-called parity-time
($\PT$) symmetry condition, $\epsilon_1=\epsilon_2^*$ or
$\alpha=\alpha_1=\alpha_2^*$ (assuming equal radii). In this case, the
dipolar analysis above suggests a divergence at the critical
separation $d_c$ corresponding to $L^3=\{|\alpha|,\sqrt{2}|\alpha|\}$,
illustrated in the bottom contour plot of \figref{fig1}(c), assuming
$|\Im\epsilon_{1,2}|=0.1$. It also follows that under finite loss
rates, the emission from gain--loss dimers can be made arbitrarily
larger than that of their passive counterparts. Note that in order to
capture the enhancement factor associated with active dimers, the
induced polarization effect (captured by our quasistatic analysis to
first order in $d/R$) must be included, emphasizing the importance of
geometry along with gain in realizing maximum ASET; the former has a
significantly smaller effect on passive dimers.

Deviations from zero-loss conditions lead to different scalings in
active versus passive dimers: for small but finite $\Im\alpha \ll
|\Re\alpha|$, the passive transfer rate, $\PhiHT \sim (\frac{\Re
  \alpha}{\Im \alpha})^2$, illustrating a significantly more dramatic
increase in flux rates with decreasing losses than is otherwise
observed in the planar geometry discussed
above~\cite{miller2015shape}. Additional enhancements arise in active
dimers. For instance, under an equally small breaking of $\PT$
symmetry in our example above, i.e. $\alpha = \alpha_1 =
\alpha^*_2+i\delta$, one finds that $\PhiHT \sim (\frac{\Im
  \alpha}{\Re \alpha})^2 (\frac{\Im\alpha}{\delta})^2$. Considering
the typically large loss rates of metals near the plasma frequency,
i.e. $\Im \alpha / \Re\alpha \sim 1$, it is clear that in practice,
one can achieve larger enhancement factors in active dimers as
compared to passive dimers. Note that although we focus here on a
$\PT$-symmetric configuration as a convenient illustration of
amplification phenomenon, similar results arise under different
scenarios, as described by the divergence condition above.

While the DA offers intuitive and analytical insights into energy
exchange in the subwavelength regime, it fails to capture many
important, finite-size effects that result from second- and
higher-order scattering artifacts, and must therefore be supplemented
by exact calculations if more quantitative predictions are
desired. Nevertheless, as shown in \figref{fig2}(b), when compared
against the SA above, with flux rates given by \eqref{PhiNF} and
\eqref{PhiFF}, the DA and exact predictions exhibit close agreement
whenever $d\gtrsim R$, suggesting that the DA is sufficient to
understand the main features of energy transfer at intermediate to
large separations. It is also evident from the DA that the ratio of
ASET to ASE, $\frac{\PhiHT}{\Phirad} \sim \frac{(R/d)^6}{(k_0 R)^3}$,
favoring absorption to radiation as $k_0 R \to 0$, as illustrated in
\figref{fig2}. Furthermore, although our dipolar analysis suggests a
unique $L$ at which $\PhiHT \to \infty$, finite geometries support
many such modes and there exists multiple critical separations and
quasistatic divergences, an example of which is shown in
\figref{fig2}(c)(d), which delineate lasing transitions and strong,
distance-dependent enhancements at $d \lesssim R$ not predicted by
DA. In particular, \figref{fig2}(c) shows the flux rates under $\PT$
symmetry, corresponding to $\Re\epsilon=-1.522$ and
$\Im\epsilon_1=-\Im\epsilon_2=-0.05$, illustrating the appearance of a
subwavelength resonance (otherwise absent at far-away separations) at
$d\approx 0.317R$ and $\omega_0 R/c \approx 0.25$ that splits into two
resonances at $d/R \approx 0.306R$, whose frequencies $\omega_0^{\pm}$
move farther apart (white dashed lines in the top contour plot) with
decreasing $d$. Such a resonant coupling mechanism results in an
ultra-large red shift $\omega_0^{-} \to 0$ of one of the branches, as
$d \to d_c$, eventually leading to the quasistatic divergence and
better illustrated in the bottom figure of \figref{fig2}(c), which
shows the spectrum corresponding to three different separations,
denoted by white dots. While the DA does not predict such a low-$d$
divergence, which arises due to higher-order scattering effects, it
does predict the right scaling of $\PhiHT/\Phirad$ with the various
parameters.

\begin{figure*}[t!]
\centering \includegraphics[width=\linewidth]{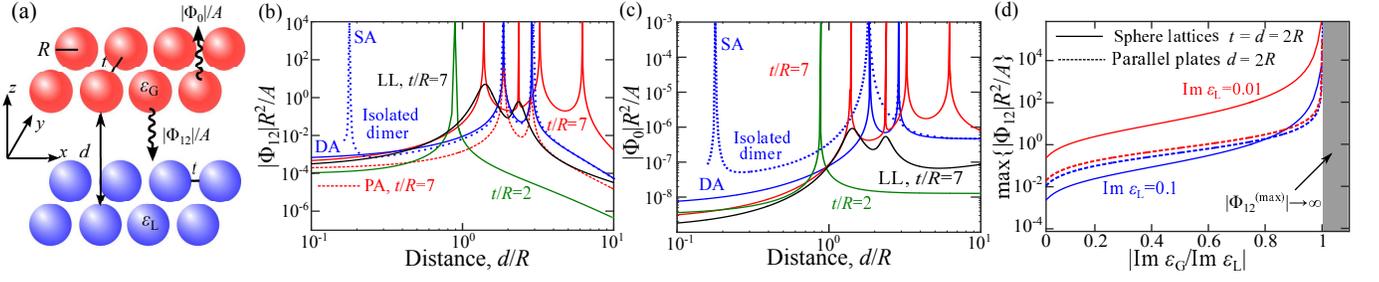}
\caption{(b) Flux transfer $|\Phi_{12}|R^2/A$ and (c) far-field flux
  $|\Phi_{0}|R^2/A$ associated with the system shown schematically in
  (a), involving an infinite, two-dimensional lattices of gain and
  loss spheres of equal radii $R$ and period $t\approx d$, and
  separated by a (varying) vertical distance $d$, for different
  choices of $t/R \gtrsim 1$ and fixed values of $\Re\epsilon=-1.95$
  and $\Im\epsilon_{1}=-\Im\epsilon_2=-0.01$. Also shown are the
  corresponding flux rates obtained using a simple pairwise
  approximation (PA, dashed red lines) that ignores multiple
  scattering (see text), or associated with either passive spheres
  (LL, black solid lines) or isolated dimers (blue lines, both DA and
  SA). The flux rates are normalized by either the dimensionless unit
  areas $A/R^2$ in the case of lattices, with $A=(t+2R)^2$, or $A=4\pi
  R^2$ in the case of an isolated dimer. (d) compares the maximum
  achievable flux rate $|\Phi_{12}|R^2/A$ in sphere lattice (solid
  lines) versus planar (dashed lines) geometries as a function of the
  ratio $\Im\epsilon_1/\Im\epsilon_2$ (relative overall permittivity
  of the gain spheres/plates) for two different choices of
  $\Im\epsilon_2=\{0.01,0.1\}$ (red, blue) and fixed lattice
  parameters $d/R=t/R=2$.  }
\label{fig:fig3}
\end{figure*}

The analysis above suggests that a proper combination of gain,
geometry, and subwavelength operating conditions can provide optimal
conditions for achieving ASET $\gg$ ASE below the LT. In what follows,
we consider a more practical and interesting, extended geometry,
involving lattices of spheres that exchange energy among one another,
where one can potentially observe even larger enhancements, leaving
open the possibility of further improvements in other
geometries~\cite{huth2010shape,incardone2014heat,narayanaswamy2008near}.
Because exact calculations of flux rates in such a structure are far
more complicated~\cite{zabkov2011plasmon}, we restrict ourselves to
quasistatic situations that lie within the scope of our DA.

%\emph{Extended structures:} 

\subsection{Sphere lattices}

The combination of reduced loss rates and resonant, near-field
enhancements potentially achievable in extended geometries could lead
to orders of magnitude larger heat flux rates compared to planar
geometries. In fact, as we showed recently
in~\citeasnoun{miller2015shape}, structures comprising tightly packed,
pairwise-additive dipolar radiators can approach the fundamental
limits of radiative energy exchange imposed by energy conservation. In
what follows, we analyze more realistic versions of such structures,
albeit under gain, demonstrating the possibility of achieving
significant and widely tunable near-field and material flux
enhancements.

We consider two vacuum-separated square lattices of gain--loss
nanospheres having equal radii $R$, lattice spacing $t$, and
surface--surface separation $d$, depicted in \figref{fig3}(a). As
noted above, the radiation between and from such structures will, to
lower order in $\{d,t\}/R$, depend on the local corrections to the
polarizabilities of each individual sphere. The generalization of the
DA to consider such a situation yields the following set of equations
for the effective polarizabilities of each sphere:
\begin{align}
&\left[\frac{1}{\alpha_{G,z}^{(0)}}-\frac{1}{(2+t/R)^3}\sum_{\substack{n_1,n_2=0 \\ n_1+n_2\neq0}}^{\infty}
    \frac{1}{(n_1^2+n_2^2)^{3/2}}\right]\alpha^\text{eff}_{G,z} \nonumber \\ &-
  \left[\frac{1}{(2+t/R)^3}\sum_{\substack{n_1,n_2=0}}^{\infty}
    \frac{n_1^2+n_2^2-2(d/t)^2}{[n_1^2+n_2^2+(d/t)^2]^{5/2}}\right]\alpha^\text{eff}_{L,z}=1 \\ &\left[\frac{1}{\alpha_{G,\parallel}^{(0)}}-\frac{1}{(2+t/R)^3}\sum_{\substack{n_1=0,n_2=0\\ n_1+n_2\neq0}}^{\infty}\frac{n_2^2-11n_1^2}{(n_1^2+
    n_2^2)^{5/2}}\right]\alpha^\text{eff}_{G,\parallel} \nonumber \\ &-
  \left[\frac{1}{(2+t/R)^3}\sum_{\substack{n_1,n_2=0}}^{\infty}
  \frac{(d/t)^2+n_2^2-11n_1^2}{[n_1^2+n_2^2+(d/t)^2]^{5/2}}\right]\alpha^\text{eff}_{L,\parallel}=1,
\end{align}
in terms of the bare polarizabilities $\alpha^{(0)}_{G,L}$ and
structure parameters. (Note that there are three additional equations,
which we have chosen to omit, obtained by letting $G\leftrightarrow L$
.)

\Figref{fig3} shows (b) $\Phi_{12}$ and (c) $\Phi_0$ in the
subwavelength regime $k_0 R = 0.01$, normalized by the dimensionless
lattice area $A/R^2 = (2+t/R)^2$, assuming spheres of
$\epsilon_{1,2}=-1.95\pm 0.01i$ and for various $t=\{2,7\}R$. To
understand the range of validity of the DA with respect to $d/R$, we
once again compare its predictions against our semi-analytical
formulas (SA) in the case of isolated dimers (dotted blue lines),
showing excellent agreement in the range $d/R > 1$; note, however, the
failure of DA to predict the additional peak at low $d/R \approx
0.2$. Restricting our analysis to large separations, one finds that
the presence of additional spheres causes significant enhancements and
modifications to the flux rates, leading to complicated, non-monotonic
dependences on geometric parameters such as $t$. To illustrate the
importance of multiple-scattering among many particles, we also show
results obtained using a simple pairwise-additive (PA) approximation
(dashed lines), in which the flux rates associated with pairs of
spheres are individually summed.

\Figref{fig3}(d) compares the performance of sphere lattices against
that of parallel plates, showing the maximum achievable
$|\Phi_{12}|/(A/R^2)$ as a function of the relative gain/loss rate
$\Im\epsilon_1/\Im\epsilon_2$ for fixed $d/R=t/R=2$ and multiple loss
rates $\Im\epsilon_2=\{0.01,0.1\}$ (red and blue lines), varying $\Re
\epsilon_{1,2}$ so as to satisfy the resonant condition (obtained and
verified numerically).  As noted above, whenever $\Im\epsilon_1 < 0$
(loss compensation), it is always possible to choose geometric
parameters under which the system undergoes lasing (gray shaded
region), though this condition can only be obtained analytically for
simple structures such as the plates or dipolar spheres above. Below
the LT, it is evident that there is significant enhancement in ASET
compared to plates, especially as the lattice system approaches the
LT. Such an enhancement depends crucially on the loss rates,
decreasing with increasing $\Im\epsilon_2$, which can be explained by
the weak, logarithmic dependence of the planar flux rates on overall
loss compensation~\cite{miller2015shape}. Note that as discussed
above, at finite $R$, the DA becomes increasingly inaccurate in the
limit $\Im\epsilon_1 \to 0$, owing to the finite skin depth
effect~\cite{miller2016fundamental,Zhang07}. Our calculations
therefore offer only a qualitative understanding of the trade-offs in
exploiting particle lattices as opposed to plates.  Under losses
$\Im\epsilon_{2} \approx 0.1$ typical of plasmonic materials, we find
that parallel plates exchange more energy compared to sphere lattices
for a wide range of gain parameters (except close to the LT), while
the latter dominate at smaller $\Im\epsilon_{2}$ and can be greatly
enhanced by the presence of even a small amount of gain. Note that
while we have chosen to investigate only the case $\{t,d\}/R=2$ in
order to ensure the validity of the DA, potentially larger
enhancements are expected to arise at shorter distances or lattice
separations, but such an analysis requires a full treatment of ASET in
these extended systems, including both finite size and nonlinear
effects~\cite{cerjan2015quantitative,pick2015multimode}. Nevertheless,
our results provide a glimpse of the opportunities for tuning ASET in
structured materials.

\begin{figure}[t!]
\centering \includegraphics[width=1\linewidth]{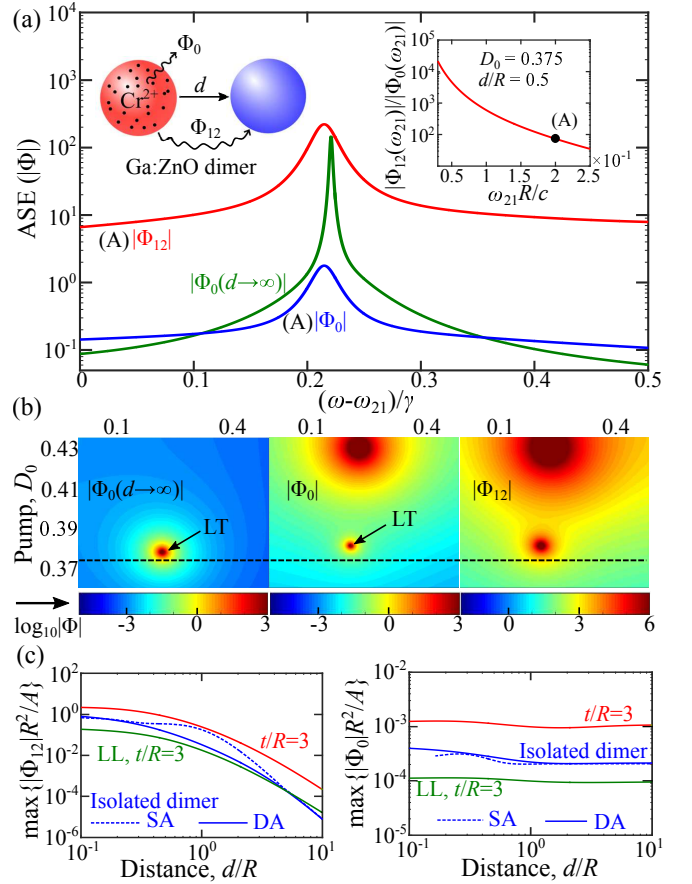}
\caption{(a) Far-field flux $\Phi_0(\omega)$ (blue line) and
  flux-transfer $\PhiHT(\omega)$ (red line) spectra of a dimer
  consisting of two Ga-doped zinc-oxide spheres of radii $R =
  0.2c/\omega_{21}$, separated by a distance $d/R=0.5$. One of the
  spheres is doped with Chromium (Cr$^{2+}$) ions having transition
  wavelength $\lambda_{21}=2.51\mu$m, and pumped to a population
  inversion $D_0 = 0.375~(\hbar\gamma_\perp/4\pi^2 g^2)$. Also shown
  is the far-field emission $\Phirad(d\to\infty)$ of the isolated gain
  sphere (green line). The top inset shows the peak ratio
  $\Phi_{12}^{\text{max}}/\Phi_0^{\text{max}}$ with respect to changes
  in $R$, keeping $d/R$ and $D_0$ fixed. (b) Contour plots
  illustrating variations in $\Phirad$ (left/middle) and $\PhiHT$
  (right) with respect to $D_0$, with the black dashed lines
  indicating operating parameters in (a). (c) Maximum spectral flux
  rates $|\Phi_{12}(\omega)|R^2/A$ (left) and $|\Phi_0(\omega)|R^2/A$
  (right) for extended sphere lattices comprising GZO gain-loss
  spheres operating at $D_0 = 0.3~(\hbar\gamma_\perp/4\pi^2 g^2)$,
  well below the LT, but of radii $R\sim 0.05c/\omega_{21}$, as a
  function of $d/R$ and for different values of $t/R$. Also shown are
  the flux rates of passive lattices (LL, black solid lines), obtained
  by letting $D_0 = 0$.}
\label{fig:fig4}
\end{figure}

\subsection{Real materials}

The ability to achieve gain at subwavelength frequencies is highly
constrained by size and material considerations. In what follows, we
describe ASET predictions in a potentially viable material
system. Consider a sphere dimer consisting of two ion-doped metallic
spheres, shown schematically on the inset of \figref{fig4}. While
there are many material candidates, including various choices of
metal-doped oxides and chalcogenides~\cite{Boltasseva13}, for
illustration, we consider a medium consisting of ($2wt\%$) Ga-doped
zinc oxide (GZO) that is further doped with 4-level Chromium
(Cr$^{2+}$) ions, in which case the transition wavelength lies in the
near infrared. The permittivity and gain profile of the ions and GZO
are well described by \eqref{gaindispersion}, with $\omega_{21} =
0.75\times 10^{15}$~rad/s, $\gamma_{\bot}\approx 0.02\omega_{21}$,
and~\cite{Boltasseva13,Lambrecht08:casimir,Stone12},
\begin{equation}
  \epsilon_r(\omega)=\epsilon_{\infty}-\frac{\omega_p^2}{
    \omega(\omega+i\Gamma_p)} + \frac{f_1\omega_1^2}{\omega_1^2
    -\omega^2 -i\omega\Gamma_1}
\label{eq:host}
\end{equation}
where $\epsilon_{\infty}=2.475$, $f_1=0.866$,
$\omega_p=2.23\omega_{21}$, $\Gamma_p=0.0345\omega_p$,
$\omega_1=9.82\omega_{21}$, and $\Gamma_1=0.006\omega_1$. These
parameters dictate dimer sizes and configurations needed to operate in
the subwavelength regime.
    
\Figref{fig4}(a) shows $\PhiHT$ (red line) and $\Phirad$ (blue line)
for one possible dimer configuration, corresponding to $R = 0.2
c/\omega_{21} \approx 80$nm, $d/R=0.5$, and population inversion
$D_0=0.375~(\hbar\gamma_\perp/4\pi g^2)$, demonstrating orders of
magnitude larger ASET compared to ASE within the gain
bandwidth. Noticeably, the emission from an isolated sphere under the
same gain parameters (green line) is significantly larger, evidence of
an increased LT due to the presence of the lossy sphere. The flux
spectra of this system are explored in \figref{fig4}(b) with respect
to changes in $D_0$, illustrating the appearance of the subwavelength
peak and large $\PhiHT \gg 1$. As expected, the LT corresponding to
the first peak occurs slightly above $\Im\epsilon_L \approx 0.37$,
which is the threshold gain needed to compensate material loss, at
which point $\Im\epsilon_1 <0$.  The black dashed lines in the
contours denote the operating parameters of \figref{fig4}(a),
confirming that the system lies below the LT. As expected, smaller
dimers lead to larger $\frac{\PhiHT}{\Phirad}\sim (k_0 R)^{-3}$, as
illustrated by the top inset of \figref{fig4}(a). \Figref{fig4}(c)
shows the flux rates (red and blue lines) corresponding to extended
lattices of spheres comprising the same GZO gain--loss profiles and
with radii $R=0.05c/\omega_{21}\approx 20$nm (in the highly
subwavelength regime), in a situation where the system is well below
the LT, which occurs at $D_0 = 0.3~(\hbar\gamma_\perp/4\pi^2
g^2)$. Noticeably, the flux rates are significantly larger than the
rates achievable in passive structures (green solid lines).
 
\section{Concluding Remarks}
Our predictions shed light on considerations needed to achieve large
ASET between structured active--passive materials, attained via a
combination of loss compensation in conjunction with near-field
effects. While our work follows closely well-known and related ideas
in the areas of near-field heat transport and nano-scale lasers
(e.g. spasers), the possibility of tuning and enhancing heat among
active bodies in the near field is only starting to be
explored~\cite{Minnich15,heatfluxFan15}. Our analysis, while
motivating and correct in regimes where ASE domiantes stimulated
emission, ignores important nonlinear and radiative-feedback effects
present in gain media as the LT is approached, nor have we considered
specific pump mechanisms which will necessarily affect power
requirements and ASET predictions~\cite{gu2013,weiliang15:purcell},
especially above threshold.  To answer such questions, future analyses
based on full solution of the Maxwell--Bloch
equations~\cite{Scully97,Siegman71} or variants
thereof~\cite{Stone12,pick2015multimode} are needed.

%One possibility is to exploit finite clusters of
%metallic nanoparticles operating in the subwavelength regime,
%following a similar approach for reaching limits to heat transfer
%between extended, passive
%structures~\cite{Rodriguez15:limits}. 

%Through such microscopic analyses, it may then be possible to fully
%examine the possibility of achieving large thermal extraction and
%cooling in active--passive systems.

\emph{Acknowledgments.---} We would like to thank Steven G. Johnson
and Zin Lin for useful discussions. This work was partially supported
by the Army Research Office through the Institute for Soldier
Nanotechnologies under Contract no. W911NF-13-D-0001, the National
Science Foundation under Grant no. DMR-1454836 and by the the
Princeton Center for Complex Materials, a MRSEC supported by NSF Grant
DMR 1420541.
%\pagebreak
%\begin{center}
%\textbf{\large APPENDIX}
%\end{center}

\begin{appendix}

\section{Vector spherical harmonics}
\label{appendix:app}

When deriving the flux rates associated with two spheres, we employed
the following spherical-vector functions:
\begin{align}
\label{eq:Mp}
  \vec{M}^{(p)+}_{\ell m}(k \vec{x}) &= z^{(p)}_\ell(k r)
  \vec{V}^{(2)}_{\ell m}(\theta,\phi), \\ \vec{M}^{(p)-}_{\ell m}(k
  \vec{x}) &= \zeta^{(p)}_\ell(k r) \vec{V}^{(3)}_{\ell
    m}(\theta,\phi) \nonumber \\ &+ \frac{z^{(p)}_\ell(k r)}{k r}
  \sqrt{\ell(\ell+1)} \vec{V}^{(1)}_{\ell m}(\theta,\phi),
\label{eq:Mm}
\end{align}
where $z^{(p)}_\ell$ are spherical Bessel ($p=1$) and Hankel ($p=3$)
functions of order $\ell$, $\zeta^{(p)}_\ell(x) =
\frac{1}{x}\frac{d}{dx} [x z^{(p)}_\ell(x)]$, and $\vec{V}^{(p)}_{\ell
  m}$ and associated spherical vector harmonics~\cite{Chew99},
\begin{align}
\vec{V}^{(1)}_{\ell m}(\theta,\phi) &= \hat{\vec{r}} Y_{\ell m}
\\ \vec{V}^{(2)}_{\ell m}(\theta,\phi) &= \frac{1}{\sqrt{\ell
    (\ell+1)}} \left(-\boldsymbol{\hat{\phi}} \frac{\partial Y_{\ell
    m}}{\partial \theta} + i\boldsymbol{\hat{\theta}}
\frac{m}{\sin\theta} Y_{\ell m} \right) \\ \vec{V}^{(3)}(\theta,\phi)
&= \frac{1}{\sqrt{\ell (\ell+1)}} \left(\boldsymbol{\hat{\theta}}
\frac{\partial Y_{\ell m}}{\partial \theta} + i\boldsymbol{\hat{\phi}}
\frac{m}{\sin\theta} Y_{\ell m} \right),
\end{align}
which satisfy the following orthogonality relations:
\begin{align*}
  \oint_S \vec{V}^{(p)}_{\ell m} \cdot \vec{V}^{(p')*}_{\ell' m'} &=
  \delta_{\ell\ell'} \delta_{p p'} \delta_{m m'} \\ \oint_S d\Omega\,
  \vec{V}^{(3)}_{\ell m} \times \vec{V}^{(2)*}_{\ell' m'} \cdot
  \vec{\hat{r}} &= -\oint_S d\Omega\, \vec{V}^{(2)}_{\ell m} \times
  \vec{V}^{(3)*}_{\ell' m'} \cdot \vec{\hat{r}} \\ &= -\delta_{\ell
    \ell'}\delta_{m m'} \\ \int_{V_i} d\vec{x}' \,
  \vec{M}^{(1)+}_{\ell m}(k\vec{x}') \cdot \vec{M}^{(1)+*}_{\ell'
    m'}&(k\vec{x}') \\ &\hspace{-1in}= R_i^2\Im \left[k_i^*
    z^{(1)}_\ell(k_i R_i) \zeta^{(1)*}_\ell(k_i R_i) \right]
  \frac{\delta_{\ell \ell'} \delta_{m m'}}{k_0^2 \Im \epsilon_i},
  \\ \int_{V_i} d\vec{x}' \, \vec{M}^{(1)-}_{\ell m}(k\vec{x}') \cdot
  \vec{M}^{(1)-*}_{\ell' m'}&(k\vec{x}') \\ &\hspace{-1in}= R_i^2\Im
  \left[k_i^* z^{(1)*}_\ell(k_i R_i) \zeta^{(1)}_\ell(k_i R_i) \right]
  \frac{\delta_{\ell \ell'} \delta_{m m'}}{k_0^2 \Im \epsilon_i}.
\end{align*}

\end{appendix}
%%%%%%%%%%%%%%%%%%%%%%%%%%%%%%%%%%%%%%%%%%%%%%%%%%%%%%%%%%%%%%%%%%%%

\bibliographystyle{unsrt}
\bibliography{photon}

\end{document}